\documentclass[3p]{elsarticle}

\usepackage[utf8]{inputenc}
\usepackage{amsmath}
\usepackage{amsfonts}
\usepackage{comment}
\usepackage{float}
\usepackage[hyphens]{url}
\usepackage[hidelinks]{hyperref}
\hypersetup{breaklinks=true}
\usepackage{cleveref}

\newcommand{\puo}{PuO$_{2}$}
\newcommand{\degree}{\ensuremath{^{\circ}}}

\title{Synthesis parameter effect detection using quantitative representations and high dimensional distribution distances}
\author{Alex Hagen, Shane Jackson}
\date{January 2023}
\journal{Chemometrics and Intelligent Laboratory Systems as a short communication}

\begin{document}

\begin{abstract}
Detection of effects of the parameters of the synthetic process on the microstructure of materials is an important, yet elusive goal of materials science. We develop a method for detecting effects based on copula theory, high dimensional distribution distances, and permutational statistics to analyze a designed experiment synthesizing plutonium oxide from Pu(III) Oxalate.  We detect effects of strike order and oxalic acid feed on the microstructure of the resulting plutonium oxide, which match the literature well.  We also detect excess bivariate effects between the pairs of acid concentration, strike order and precipitation temperature.
\end{abstract}

\maketitle

Detecting changes in a material from differing constituents, processing steps, processing environments, and other material history is a core capability needed in material science.  Collectively called "effect detection", this process is the first step for new material discovery \cite{Belgasam2018}, material property optimization \cite{Belgasam2017}, signature discovery for material provenance discovery \cite{Burr2021}, and other generalized material property studies.  In this work, we apply high dimensional distribution statistics to the challenge of detecting effects in materials where the response variable is microstructure. To do so, we apply convolutional neural networks to micrographs of synthesized particles to generate quantitative descriptions of the imaged microstructure, and then perform high-dimensional and statistically rigorous effect detection. This extension of effect detection from macroscopic properties to microstructure will have impact across material science, from the analysis of results of designed experiments to data driven provenance analysis, and towards automated material discovery.

The literature for detection of effects in materials science is deep, albeit
dominated by variants of the analysis of variance (ANOVA) methodology \cite{Moran1918}.  These methods require two limiting constraints: that the response variable must be single dimensional, and that the underlying distributions be well approximated by Gaussian distributions.  The microstructure of materials, for example in analysis of the processes generating actinides \cite{Keegan2014}, is of academic and practical interest. The detection of effects in such cases is complicated in that microstructure is a qualitative feature of a particle. Commonly, the microstructure of a particle is assessed using some sort of microscopy, such as secondary electron scanning electron microscopy (SEM).  Several have tried to taxonomize such microstructure in order to fit it into a quantitative scheme \cite{Tamasi2016}.  Borrowing from the broad field of computer vision on natural imagery, others have used convolutional neural networks to quantitatively describe the SEM micrographs \cite{Girard2021, Nizinski2022, Ly2019}.  We utilize this last technique due to its repeatability and speed.  

To extend a quantitative description scheme into effect detection, we require a statistical test.  In the quantitative description schemes of such micrographs, the output description is high dimensional. Therefore, we require an independence test for two high dimensional variables. In our review and empirical testing, explicit tests for high dimensional dependence were unsatisfactory - for example the distance correlation and covariance \cite{Szekely2013} has difficulty detecting non-linear or non-monotone relationship \cite{Remillard2009, Kosorok2009}, and in our experiments either returned pathological results\footnote{such as correlations provably smaller than that of two random variables, in fact truly negative}, or in the copula formulation \cite{Remillard2009} had many false positives\footnote{We define a false positive as the detection of an effect between two truly random vectors or matrices}. Therefore, we turn to novel methods to detect high dimensional effects. 

In one dimension statistical distances can be readily defined; however in higher dimensions, the ambiguity in defining the cumulative distribution function permits many possible definitions of distance\footnote{In the two dimensional case, the CDF can be defined as $F=P\left(X<x,Y<y\right)$ or $F\left(X<x, Y>y\right)$ or any permutation thereof.}. Furthermore, due to their computational cost, many methods rely on assumptions about the underlying distributions which can be problematic if the \emph{a priori} knowledge of the dataset is limited. For example, an analogue of our method is the Frechet Inception Distance (FID) \cite{Heusel2017}, often used to compare synthetic images to real images in the natural imagery literature.  Underlying that distance is the assumption that all distributions are Gaussian and therefore can be compared by comparing only moments; an assumption which can be easily disproven for many materials science related datasets.  Non-parametric high  dimensional distances such as the d-dimensional Kolmogorov-Smirnov distance (ddks) \cite{Hagen2021} and the Mean Maximum Discrepancy (mmd) \cite{Gretton2012} are then the focus of our study. 

We investigate a set of micrographs of ex-Pu(III) oxalates calcined to \puo{}, generated in a 76 run designed experiment \cite{Hainje2022}. The designed experiment varied the values of 7 parameters: (A) Oxalic Acid Feed, (B) Addition time, (C) Digestion time, (D) HNO$_{3}$ concentration, (E) Strike Order, (F) Precipitation temperature, and (G) Pu concentration. Their possible values are listed in \cref{table:enc}.  Of the 76 runs, 50 met quality control requirements and were investigated in this study. We prepocessed the data by first isolating each particle within an overall micrograph, and then encoding it using a vector quantizing variational autoencoder (VQVAE) from \cite{Girard2021}. %This data is currently available only to United States National Nuclear Security Administration Defense Nuclear Nonproliferation and related projects.

\begin{table}
\begin{center}
\setcounter{table}{0}
    \caption{Pu(III) oxalate precipitation synthetic parameters and their respective label encodings. Reproduced from \cite{Hainje2022} with permission.}
    \label{table:enc}
    \vspace{1ex}
\begin{tabular}{|c c | c c c|} 
 \hline
 & Parameter & Encoding: 0 & 1 & 2 \\ [0.5ex] 
 \hline\hline
 A & Oxalic acid feed & 0.9M solution & solid & - \\ 
 %\hline
 B & Addition time (min.) & 0 & 20 & 40 \\
 %\hline
 C & Digestion time (min.) & 40 & 20 & 0 \\
 %\hline
 D & HNO$_3$ concentration (M) & 1.0 & 2.0 & 3.0 \\
 %\hline
 E & Strike order & reverse & direct & - \\
 %\hline
 F & Precipitation temperature ($\degree{}$C) & 30 & 50 & - \\
 %\hline
 G & Pu concentration (g/L) &  10 & 30 & 50 \\ [0.5ex] 
 \hline
\end{tabular}
\end{center}
\end{table}

We have developed a distribution distance and perturbation based method for detecting the effect of a parameter on the high dimensional response based on the probability integral transform.  Given $N$ different particles with input parameters $\vec{x} = \left[x_{0}, x_{1}, \dots, x_{N}\right]^{T}$ and response $\vec{y} = \left[y_{0}, y_{1}, \dots, y_{N}\right]^{T}$ where the elements of $x \in \mathbb{R}^{d_x}$ and $y \in \mathbb{R}^{d_y}$.  

In the case of effect detection, our objective is to identify and quantify the probability distribution of the input parameters, $\mathcal{X}$, and the response variables $\mathcal{Y}$.  To do so, we consider the probability distribution of the joined parameter $\vec{z} = \left(\vec{x},\vec{y}\right)$ we denote as $\mathcal{Z}$.  By permuting the rows of $\vec{x}$ while keeping $\vec{y}$ fixed, a second joined parameter $\vec{\tilde{Z}}$ is generated which has the property of conditional independence.  Comparing $\vec{Z}$ and $\vec{\tilde{Z}}$ is done via two-sample tests such as the $d$-dimensional Kolmogorov-Smirnov (ddKS) test \cite{Hagen2021}, or the maximum mean discrepancy (MMD) test \cite{Gretton2006}.

In practice, we can construct one data matrix with dimension $N\times\left(d_{x} + d_{y}\right)$ by concatenating samples from $\vec{x}$ and $\vec{y}$ into data matrices $\mathbf{X}$ and $\mathbf{Y}$, denoting this concatenation $\mathbf{Z}$. We generate samples from the joint distribution by randomly selecting rows from the data matrix. We generate a sample from the joint independent distribution by sampling rows from $\mathbf{Z}$ and randomly permuting only the $\mathbf{X}$ rows, or equivalently permuting the entries of the column of the data matrix associated with $\mathbf{X}$. We denote this permuted data matrix as $\tilde{\mathbf{Z}}\equiv \left(\pi\left(\vec{x}\right),\vec{y}\right)$, with $\pi\left(\cdot\right)$ a random row-wise permutation.

To determine whether $\mathbf{Z}$ and $\tilde{\mathbf{Z}}$ are drawn from the same distribution, we measure their distribution distance with the aforementioned distances.  These distances are, in general, only constrained to be a positive real number and the statistical significance of a given distance may be unknown. We use the permutation test \cite{Fisher1935} to compute exact statistical significance of a given distance. This significance effectively becomes the measure of effect detection: if the parameter and response are independent, the significance is below threshold $1 - \alpha$ with confidence $1 - \alpha$, otherwise the response is dependent on the parameter.

The above method for effect detection admits extension to multiple synthesis parameters. Naive extension would compare the joint distribution of $\vec{x}_{1}$, $\vec{x}_{2}$ and $\vec{y}$ to the randomly shuffled combined set $\tilde{\mathbf{Z}}$. This extension would result in a measurement of the effect of $\vec{x}_{1}$ \textbf{or} $\vec{x}_{2}$ (where the or is inclusive). Instead, materials scientists are more interested in the effects of $\vec{x}_{1}$ \textbf{and} $\vec{x}_{2}$ that cannot be explained by either parameter individually.

We first compute the empirical cumulative distribution functions of $\vec{x}_{1}$ and $\vec{x}_{2}$ and used the inverse probability transform to transform them into uniformly distributed variables, $U_{1}$, and $U_{2}$.  Then, the distribution distance between $\mathbf{Z} \equiv \left(U_{1}, U_{2}, \vec{y}\right)$ and the permutation $\tilde{\mathbf{Z}} \equiv \left(\pi\left(U_{1}, U_{2}\right), \vec{y}\right)$ is a measure of the combined effect of $U_{1}$ and $U_{2}$.  To determine the sampling distribution of this distance, we can also compute the permutation test when $U_{1}$ and $U_{2}$ are separately permuted. We denote these distances $d$, $\tilde{d}_{1}$, and $\tilde{d}_{2}$.

%We first compute the copula between $\vec{x}_{1}$ and $\vec{x}_{2}$ by computing their empirical cumulative distribution functions and using that to transform them into uniformly distributed variables, $U_{1}$, and $U_{2}$.  Then, the distribution distance between $\mathbf{Z}=\left(U_{1}, U_{2}, \vec{y}\right)$ and the permutation $\tilde{\mathbf{Z}}=\left(\pi\left(U_{1}, U_{2}\right), \vec{y}\right)$ is a measure of the combined effect of $U_{1}$ and $U_{2}$.  To determine the sampling distribution of this distance, we can also compute the permutation test when $U_{1}$ and $U_{2}$ are separately permuted. We denote these distances $d$, $\tilde{d}_{1}$, and $\tilde{d}_{2}$.

The above distances are susceptible to false positives if $\vec{x}_{1}$ and $\vec{x}_{2}$ are highly correlated. To mitigate this effect, we first compute the complement of the significance of the distribution distance difference between $\vec{x}_{1}$ and $\vec{x}_{2}$.  This value is inversely proportional to the correlation between $\vec{x}_{1}$ and $\vec{x}_{2}$, and we call it $r$.  We can perform the permutation test for $r$ in two different ways, by permuting $\vec{x}_{1}$ and by permuting $\vec{x}_{2}$, denoted $\tilde{r}_{1}$ and $\tilde{r}_{2}$, respectively.

Comparing the product of $r$ and $d$ to those products of $\tilde{r}_{1} \cdot \tilde{d}_{1}$ and $\tilde{r}_{2} \cdot \tilde{d}_{2}$ provides a direct comparison of the test statistic $r\cdot d$ to its sampling distribution.  A test statistic can be constructed comparing $r \cdot d$ to $\tilde{r}\cdot\tilde{d}=\mathrm{max}\left(\tilde{r}_{1}\cdot\tilde{d}_{1}, \tilde{r}_{2}\cdot\tilde{d}_{2}\right)$.  The the significance of a multiparameter effect is then shown by the percentage of randomly sampled pairs where $r\cdot d > \tilde{r}\cdot\tilde{d}$.

\begin{figure*}
    \centering
    \includegraphics[width=0.45\textwidth]{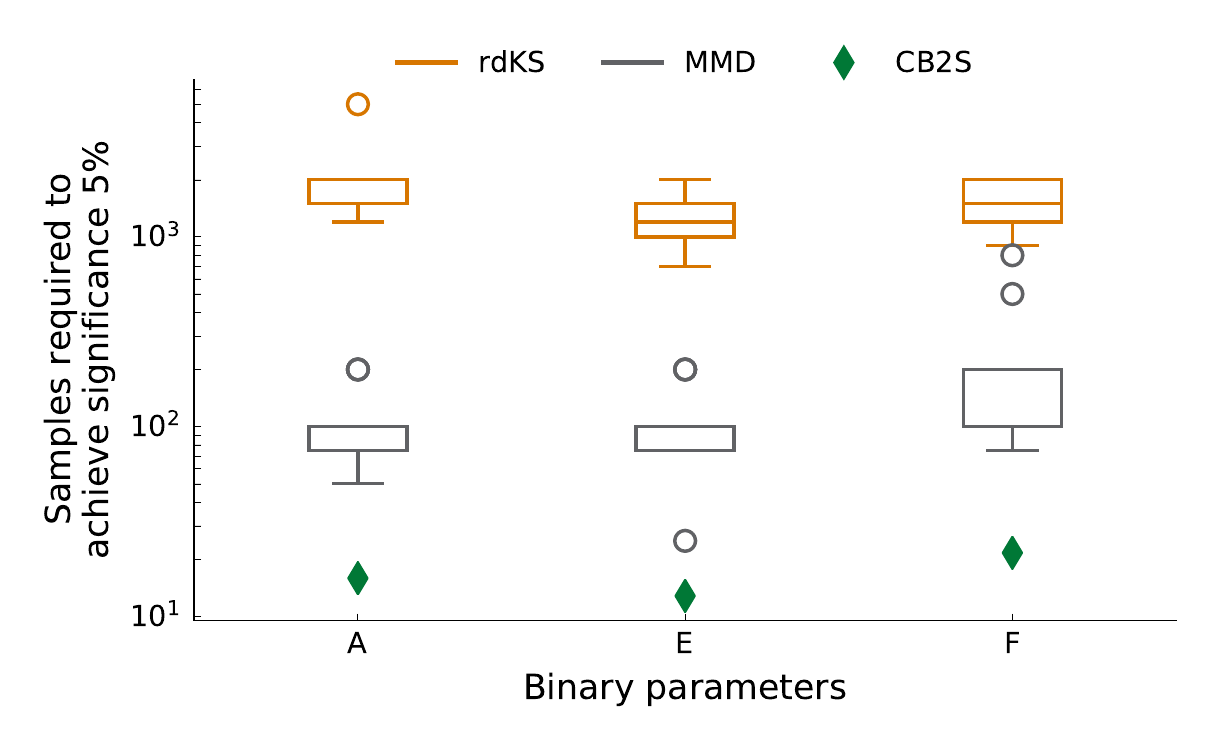} \hfill \includegraphics[width=0.45\textwidth]{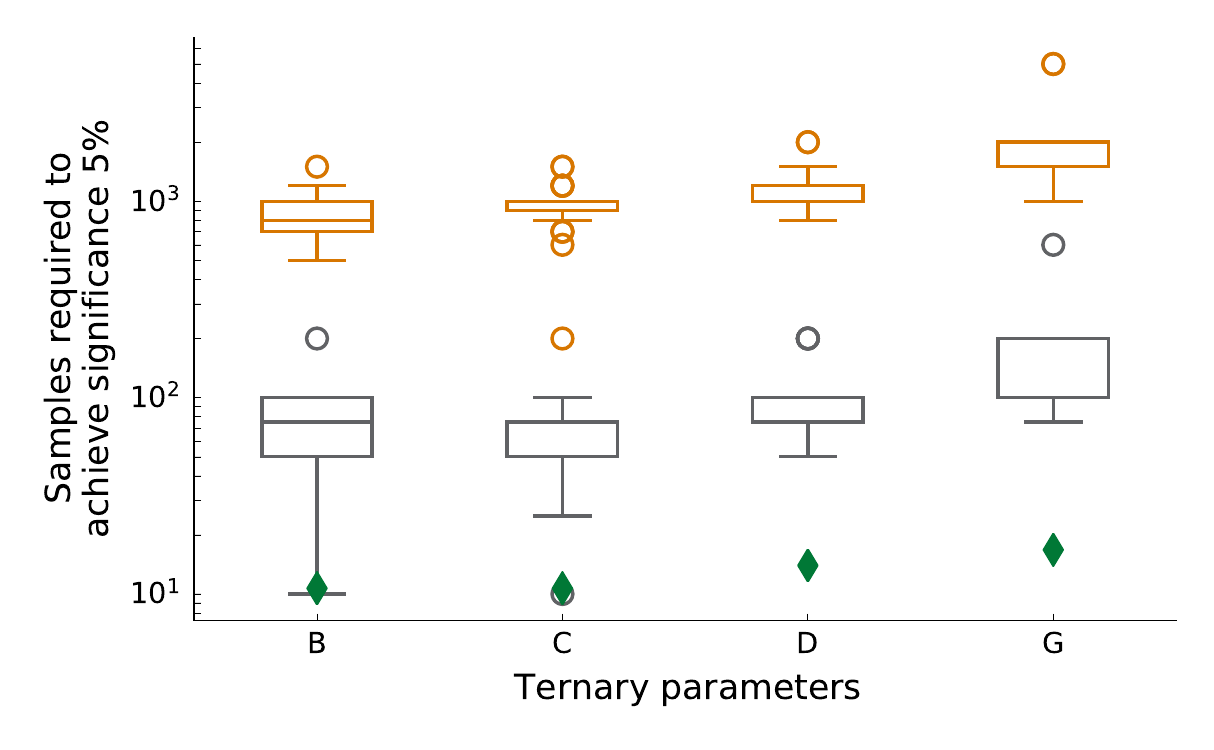}
    \\
    \includegraphics[width=0.45\textwidth]{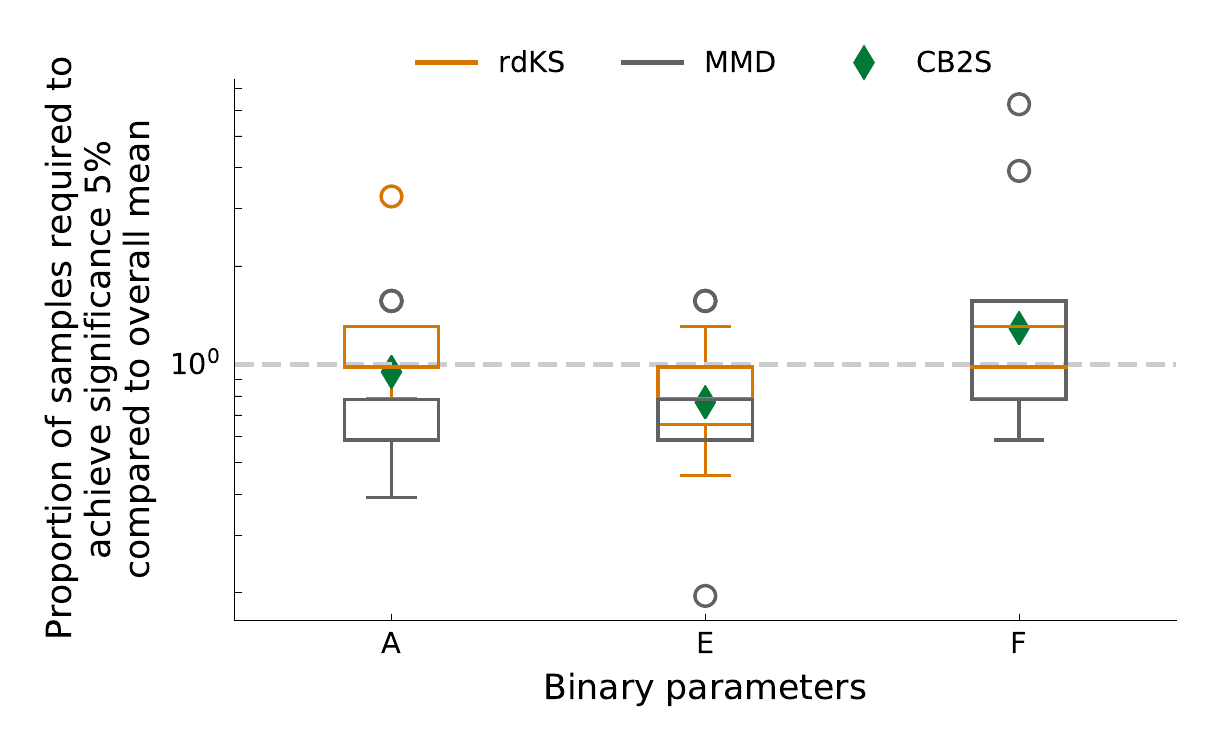} \hfill \includegraphics[width=0.45\textwidth]{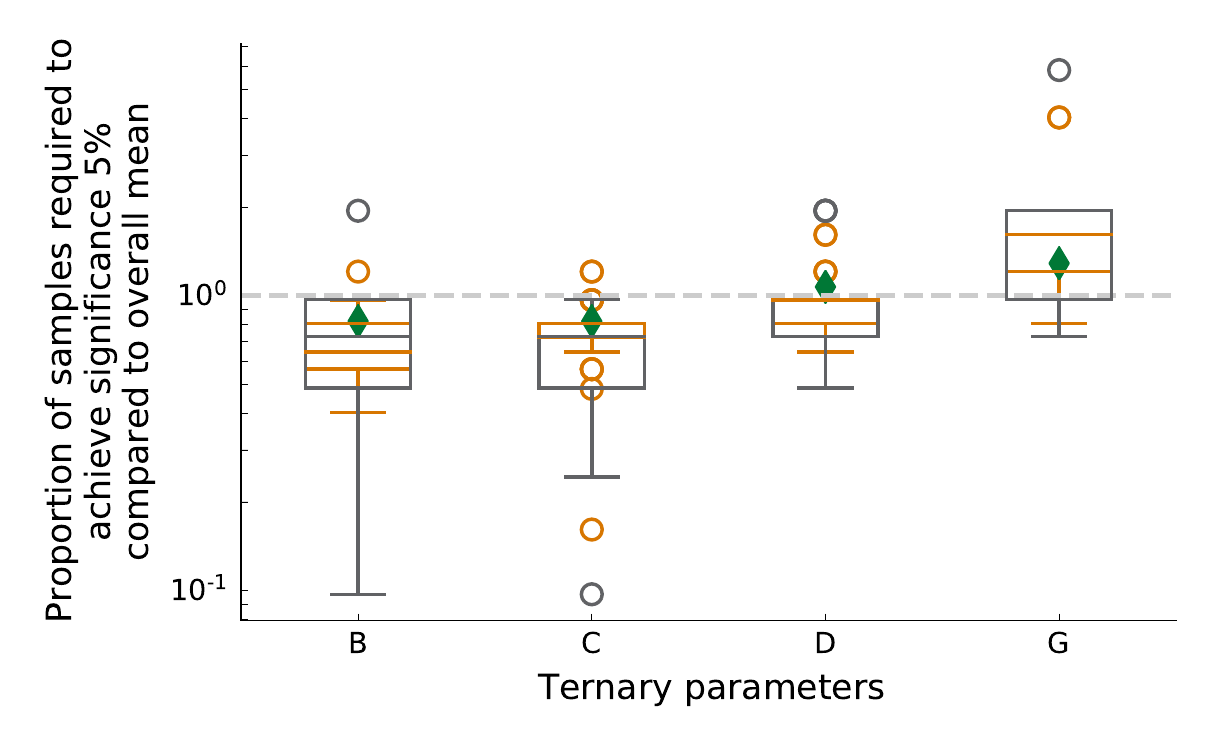}
    \\
    \caption{(First Row): Number of encoded micrographs required to reach 95\% significance when using our method and distribution distances ddKS, MMD, and a classifier based two sample test from \cite{Hainje2022} (CB2S)\\
    (Second Row): Above computations divided by the method-wise mean number of samples required to reach 95\% significance - showing that the trends detected by ddKS, MMD, and CB2S are equivalent, though their statistical power differs.}
    \label{fig:boxplots}
\end{figure*}

We performed the above method using two different distribution distances MMD and ddKS. We used sample sizes varying from 10 micrographs per computation to 5000 micrographs per computation.  We performed the permutation for each computation 100 times, and repeated all computations 25 times to account for random chance.  The results of this are shown in \cref{fig:boxplots}.

An investigation into the effect of each parameter has been previously performed in \cite{Hainje2022}. A classifier was used to categorize the micrograph of each particle into the different values for each property.  The highest achieved accuracy for each parameter was 94.1\%, 91.4\%, 91.4\%, 84.9\%, 98.0\%, 88.8\%, and 80.6\%, respectively.  Utilizing the framework of \cite{LopezPaz2016}, we can extend the classifier from \cite{Hainje2022} into a two sample test and determine the number of samples required to achieve significance $\alpha$. This allows for direct comparison of \cite{Hainje2022} to ddKS and MMD. We derive the number required in \cref{sec:cb2s} and plot the results of that computation on \cref{fig:boxplots}.

The trends displayed by all metrics on \cref{fig:boxplots} are consistent.  Among the binary parameters, an effect on microstructure by parameter E (strike order) is detected using the least number of samples, followed by parameter A (oxalic acid feed), and finally F (precipitation temperature).  Among the ternary parameters, the increasing ranked order is B/C (Addition/Digestion time - complementary variables), D (HNO$_{3}$ concentration), and finally G (Pu concentration).

\begin{figure}[t]
    \centering
    \includegraphics[width=3in]{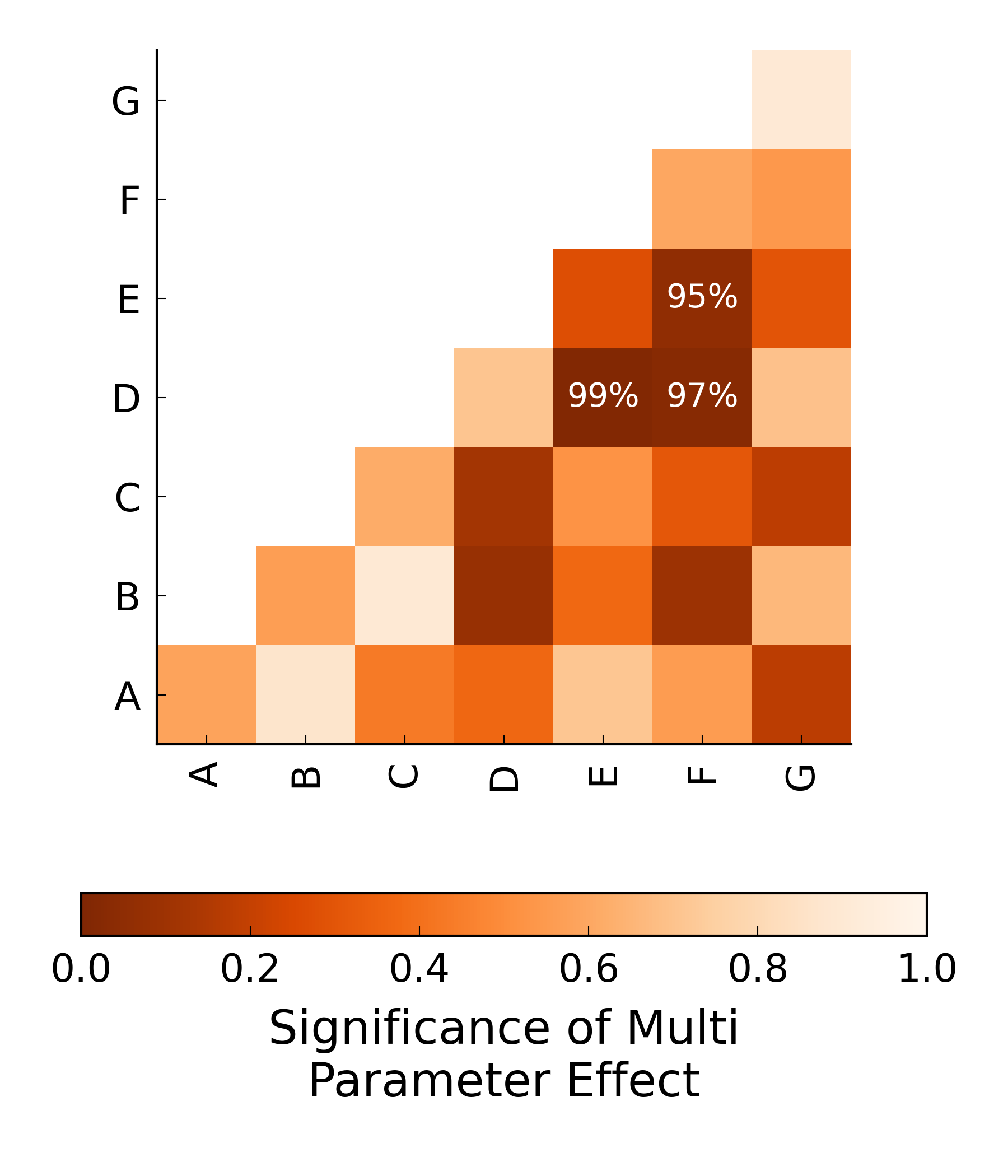}
    \caption{Significance of difference for bivariate effects between all unique pairs of variables in the dataset.  The pairs D (HNO$_{3}$ concentration) and E (Strike order) and D and F (Precipitation temperature), and E and F reach 5\% significance.}
    \label{fig:multi_parameter_effects}
\end{figure}

The multiple parameter effect method was used to investigate the possibility of bivariate effects on morphology.  Sample sizes of 5000 were chosen to ensure enough statistical power to detect second order effects.  We performed this investigation with only ddKS, as MMD requires prohibitive memory at this sample size, and the CB2S does not apply to multiparameter cases.  We performed 500 trials of comparisons between all pairs of variables. We show these results in \cref{fig:multi_parameter_effects}. Significant multiparameter effects exist between acid conentration and precipitation temperature, between acid concentration and strike order, and between strike order and precipitation temperature.  While these effects are physically plausible, we possess no other validation mechanism until more experiments are performed.

As shown by these results, this analysis using our methods are entirely consistent with those of \cite{Hainje2022}, and the materials science related discussion therefrom is more extensive than in this work.  The validation of those results with different statistical methods (albeit the same image analysis network) and that fact jointly adds credence to our methods and their results.

Our work explored several methods for effect detection in the \emph{general} sense, without the assumption of one dimensional parameters nor responses.  We showed that the distance correlation measure of independence between two parameters has undesired properties (such as negative distances in its original application, or false positives in its copula formulation).  We developed a method to avoid these difficulties, and achieved results with two different underlying distance metrics.

We extended our effect detection method to multiple parameters, and investigated bivariate effects within the dataset.  We generalized the effect size detection method using high dimensional distance metrics into one which can detect only the excess effect size due to a pair of parameters. We showed significant excess bivariate effects in the acid conentration, strike order, and precipitation temperature pairs.  Further investigation into the physics and materials science of those effects is underway.

While not applicable to the multiple parameter case, we noted several aspects that make classifier based two sample tests preferable for the single variate case. As shown in \cref{fig:boxplots}, CB2S are more powerful than distance based methods, detecting the same effect with decades fewer samples.  Further, CB2S are not dependent on linear or even monotonic effects.  The ddKS method presented was also compatible with nonlinear, nonmonotonic effects, but the MMD based method proved not to be by failing to detect the effect of an arbitrarily numbered "run" number associated with each unique set of synthesis parameters.  Finally, the CB2S is able to achieve more of the desiderata of effect detection than the distance based methods. In the general case, effect size describes a complex relationship including the extent of microstructure change and the proportion of particles which exhibit that change. The distance based methods report only the distance between the two distributions and therefore do not provide a means to separate the two possible effect manifestations.  The CB2S allows for querying of single particles and therefore can separate both manifestations.

While experimental investigation into the effects detected in this work makes up the bulk of the possible extensions, there are several improvements to the methods develop here that would be fruitful. The ddKS and MMD distance metrics are both computationally intensive, so alternative distances could provide a decrease in computational cost. Additionally, the multiple parameter effect method generated here is super-polynomial in its time complexity, and will become prohibitive in both computational cost and sample size - an alternative method to reduce those limitations would be fruitful. 

\Urlmuskip=0mu plus 1mu
\bibliographystyle{elsarticle-num} 
\bibliography{refs.bib}

\begin{appendix}

\section{Classifier Based Two Sample Test}\label{sec:cb2s}

We can cast the classifier two sample test as a draw of $N$ realizations from a binomial distribution with proportion $\lambda$.  For the number of unique values $N_{v}$, the probability of getting a correct prediction must be no higher than random chance, i.e. the proportion $\lambda$ must equal $\frac{1}{N_{v}}$.  We can compute a realization of that proportion by measuring the balanced accuracy, $a$. Therefore, if the accuracy is significantly different than $\frac{1}{N_{v}}$, then an effect has been predicted.  We can determine how many samples $N$ it takes to achieve significance using these facts. We start with Wilson's approximation to the confidence interval of a binomial proportion \cite{Wilson1927}, we can write
\begin{equation}
    p\approx \frac{aN + \frac{1}{2}z^{2}}{N+z^{2}} \pm \frac{z}{N + z^{2}}\sqrt{a\left(1-a\right)N + \frac{z^{2}}{4}},
\end{equation}
with $z$ the quantile from a standard normal distribution corresponding to the level of significance $\alpha$ (i.e. $z=1.96$ for 95\%).  When an effect exists, the upper CI of a proportion from random chance $p_{0}$ will be less than the lower limit of the CI for the classifier, $p_{a}$.  We can then solve for $N$ to find that we achieve significance at
\begin{equation}
    N = z^{2}\frac{\left[-\left(N_{v}a\right)^{2} + 2N_{v}^{2}a - 2N_{v}a + 2N_{v} - 1\right]}{\left(N_{v}a\right)^{2} - 2N_{v}a + 1}.
\end{equation}

\end{appendix}

\end{document}